\documentclass[12pt]{article}

\usepackage[a4paper]{geometry}
\usepackage[parfill]{parskip}    
\usepackage[usenames,dvipsnames]{xcolor}
\usepackage[english]{babel}
\usepackage{cite,enumerate,enumitem,booktabs,float,graphicx}

\makeatletter
\g@addto@macro\bfseries{\boldmath}
\makeatother

\usepackage{amsmath,amssymb,bm,mathtools,tensor}

\usepackage{authblk}

\usepackage[pdftex]{hyperref}
\definecolor{dark-blue}{rgb}{0.15,0.15,0.4}
\hypersetup{
  colorlinks,
  linkcolor={dark-blue},
  citecolor={blue}
}

\newcommand{\beq}{\begin{equation}}
\newcommand{\eeq}{\end{equation}}

\DeclareMathOperator{\sech}{sech}

\title{\textbf{The Kerr/CFT energy gap}}
\date{}

\author[]{Robert Penna\footnote{pennar@sunypoly.edu}}
\affil[]{
  Department of Physics \\
  SUNY Polytechnic Institute \\
  Utica, NY 13502 USA
}

\begin{document}

\maketitle

\thispagestyle{empty}

\begin{abstract}

Noether charges in gravity are always defined with respect to a reference metric, but the near-horizon extremal Kerr (NHEK) solution does not come with an obvious choice of reference metric (it is not asymptotically flat, for example).  We discuss a one parameter generalization of the NHEK solution due to Kinnersley and Kelley which interpolates between NHEK and an analytically continued version of Minkowski spacetime.  We use this interpolating family of solutions to compute the conformal energy of NHEK above the Minkowski reference metric.  
We find an interesting order of limits issue when computing the conformal energy of the Minkowski vacuum and we discuss a few natural prescriptions for fixing it, including one motivated by the Kerr/CFT correspondence.

\end{abstract}

\newpage

The throat of the extreme Kerr black hole has an emergent conformal symmetry.  This suggests that extreme Kerr might be holographically dual to a conformal field theory \cite{Bardeen:1999px,Guica:2008mu,Compere:2012jk}.  One of the difficulties in trying to understand the conjectural Kerr/CFT duality is that it is not clear how to compute the conformal energy of the throat.  The problem is that the near-horizon extremal Kerr (NHEK) metric does not come with an obvious choice of reference metric.  The NHEK metric is not asymptotically flat, so it cannot be compared to Minkowski spacetime at asymptotic infinity.  And the NHEK metric only depends on a single parameter, $J$, and for no value of this parameter does the metric reduce to Minkowski spacetime or any other reasonable reference metric.

In this work, we describe a new approach to this problem.  The basic idea can be illustrated using the Schwarzschild metric.  A standard way to compute the mass of the Schwarzschild black hole is to use the fact that it reduces to the Minkowski metric in the limit $M\rightarrow 0$.  This means we can compute its Noether charges by defining the infinitesimal charge difference between nearby Schwarzschild solutions with parameters $M$ and $M+\delta M$, and then integrating along a path through solution space that starts at $M=0$ and goes up to any desired $M$.

Now suppose that for some reason we were not allowed to vary $M$.  What else could we do?  One thing we could do is use the Zipoy-Voorhees metric \cite{Zipoy:1966btu,Voorhees:1970ywo,Stephani:2003tm,Gibbons:2017jzk}.  The Zipoy-Voorhees metric is a one parameter generalization of the Schwarzschild metric.  The new parameter, $\lambda$, is such that the Zipoy-Voorhees metric reduces to the Minkowski metric when $\lambda=0$ and it reduces to the Schwarzschild metric when $\lambda=1$.  So another way to compute the mass of the Schwarzschild black hole is to compute the infinitesimal difference between neighboring Zipoy-Voorhees metrics with parameters $\lambda$ and $\lambda+\delta \lambda$, and then integrate through solution space from $\lambda= 0$ to $\lambda = 1$.  It is a straightforward exercise to show that this gives the same answer as the standard calculation described in the previous paragraph. 

Our strategy for NHEK is to find the analogue of the Zipoy-Voorhees metric.  In fact, we do not have to look any further than the work of Kinnersley and Kelley \cite{KK1974}.  They found the generalization of NHEK that we need long ago.  Here it is:
\beq \label{eq:weyl}
ds^2 = f ( d\phi + \omega dt )^2 
		+ f^{-1} \left( e^{2\gamma} \left[ dr^2 + r^2 \sech^2 u \, du^2 \right] - \rho^2 dt^2 \right) ,
\eeq
where
\begin{align}
f		&= 4J \sech(2pu) \,, \\
\omega	&= p r \,, \\
\rho 		&= 2Jr \sech u \,, \\
e^{2\gamma} &= 4J^2 \left( \frac{\sech u}{r} \right)^{2p^2} \label{eq:gamma} \,.
\end{align}
We have simplified their expression for the metric by introducing a stereographic coordinate, $u$, which is related to the usual $\theta$ coordinate by $e^u = \cot(\theta/2)$.  The above metric is an exact solution of the vacuum Einstein equations for any real $p$.  It reduces to NHEK when $p=1$ and it reduces to a rescaled, analytically continued version of Minkowski when $p=0$.  We take the $p=0$ metric to be the reference metric for computing Noether charges. 

We compute Noether charges using the Barnich-Brandt formalism \cite{Barnich:2001jy,Barnich:2007bf}.  Let $\zeta$ be the symmetry generator and let $g_{\mu\nu}$ and $g_{\mu} + h_{\mu\nu}$ be neighboring geometries.  The infinitesimal charge difference is
\beq
\delta Q_\zeta [g] = \frac{1}{8\pi } \int_{\partial \Sigma} k_\zeta[h,g] \,, 
\eeq
where
\begin{align}
k_\zeta[h,g] =	&-\frac{1}{4} \epsilon_{\alpha\beta \mu\nu} [
			\zeta^\nu D^\mu h - \zeta^\nu D_\sigma h^{\mu\sigma} + \zeta_\sigma D^\nu h^{\mu\sigma}
					+ \frac{1}{2} h D^\nu \zeta^\mu		\notag \\
			&- h^{\nu\sigma} D_\sigma \zeta^\mu + \frac{1}{2} h^{\sigma \nu} 
					\left(  D^\mu \zeta_\sigma + D_\sigma \zeta^\mu  \right) ] dx^\alpha \wedge dx^\beta \,.
\end{align}
The integral is over the boundary of a spatial slice.  Indices are raised and lowered and the covariant derivative, $D$, is computed using $g_{\mu\nu}$.

In our application, $g_{\mu\nu}$ is given by eqns. \eqref{eq:weyl}--\eqref{eq:gamma} and $h_{\mu\nu} = \partial_p g_{\mu\nu}$.  We integrate $\delta Q$ through solution space from $p=0$ to $p=1$ holding $J$ fixed.   According to Kerr/CFT, the conformal energy, $Q_0$, is the Noether charge for $\zeta = -\partial_\phi$ \cite{Guica:2008mu}.  

\begin{figure}
\centering
\includegraphics[width=0.55\linewidth]{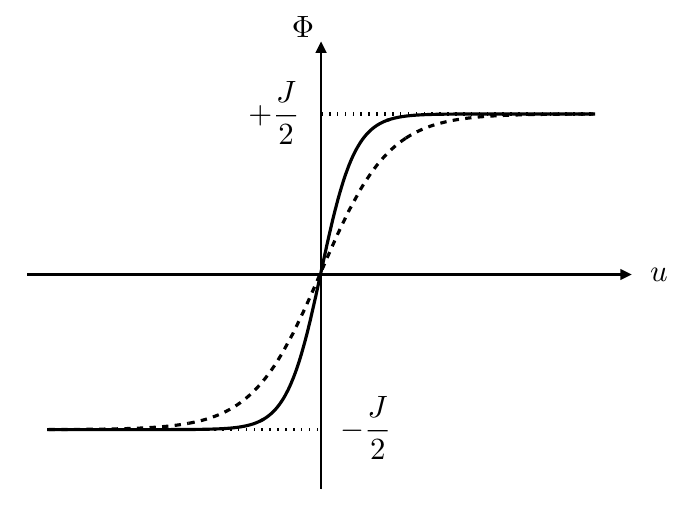}
\caption{The function $\Phi(p,u)$ for $p=1$ (solid) and $p=1/2$ (dashed).}
\label{fig:Phi}
\end{figure}

The conformal energy so defined turns out to be remarkably simple.  We find
\begin{align}
Q_0 &= \Phi( p = 1 , u = \infty ) - \Phi( p = 1 , u = -\infty)	\notag \\
	&\qquad	- \left[ \Phi(p = 0 , u = \infty ) - \Phi( p = 0 , u = -\infty ) \right] , \label{eq:Q0}
\end{align}
where
\beq
\Phi(p,u) \equiv \frac{J}{2} \tanh (2pu) \,.	\label{eq:Phi}
\eeq
The function $\Phi$ is plotted in Figure \ref{fig:Phi}.  For $p>0$, it looks like a kink separating a state with conformal energy $J/2$ at $u=+\infty$ from a state with conformal energy $-J/2$ at $u=-\infty$.  The parameter, $p$, is the inverse width of the kink.  The solutions with $p<0$ are anti-kinks.

The first line in eqn. \eqref{eq:Q0} is the contribution from the NHEK metric.  This evaluates to $J$. 

The second line of eqn. \eqref{eq:Q0} is more interesting.  This is the contribution from the Minkowski vacuum.  For small $p$, the kink becomes infinitely wide and the result depends on whether we set $p=0$ before or after setting $u=\pm \infty$.  
We have no reason to single out one choice over any other, so we content ourselves with a discussion of some of the possibilities.

One option is to set $p=0$ first and then take $u\rightarrow \pm \infty$.  In this case, the contribution from the Minkowski vacuum evaluates to zero and the energy gap between NHEK and Minkowski is $Q_0 = J - 0 = J$.   This seems like a physically reasonable choice because it is very natural to set the charges of Minkowski spacetime to zero.

Another option is to set $u = \pm \infty$ first and then set $p=0$.  In this case, the second line of \eqref{eq:Q0} evaluates to $-J$ and the energy gap between NHEK and Minkowski is $Q_0 = J - J = 0$.  Mathematically, this option has some appeal because it means all of the solutions in the Kinnersley-Kelly family with fixed $J$ sit at the same conformal energy.  The option in the previous paragraph requires a discontinuous jump in the conformal energy at $p=0$.

Finally, we consider an unusual option which nevertheless might be related to the Kerr/CFT correspondence.  This option is to set $p=0$ before $u = -\infty$ in one term, while setting $u=\infty$ before $p=0$ in the other term.   In this case, the energy gap of NHEK above the Minkowski vacuum is given by eqn. \eqref{eq:Q0} to be $Q_0 = J - J/2 = J/2$.  The reason this choice is intriguing is that the central charge of NHEK is known to be $c=12 J$ \cite{Guica:2008mu}.  Plugging $c=12J$ and $h=J/2$ into Cardy's formula in the microcanonical ensemble we obtain 
\beq
S = 2\pi \sqrt{\frac{ch}{6}} = 2\pi J \,,
\eeq
which is the Bekenstein-Hawking entropy of the extreme Kerr black hole.

Ultimately, the correct answer might depend on the question being asked.  An observer with an infinitely big laboratory would set $u=\pm \infty$ first.  An observer who only has access to a finite laboratory would set $p=0$ first.  The third option discussed above would be relevant for an observer who for some reason had access to the state at $u=+\infty$ but not the state at $u=-\infty$.

\bibliographystyle{JHEP}
\bibliography{gap}

\end{document}